\documentclass[12pt]{JHEP3}

\usepackage{amsmath,amssymb,amsfonts}
\usepackage[dvips]{graphicx}

\newcommand{\be}{\begin{equation}}
\newcommand{\ee}{\end{equation}}
\newcommand{\ba}{\begin{eqnarray}}
\newcommand{\ea}{\end{eqnarray}}
\newcommand{\bea}{\begin{array}}
\newcommand{\eea}{\end{array}}

\newcommand{\CL}{{\cal L}}

\makeatletter \@addtoreset{equation}{section} \makeatother

\preprint{KIAS-P08031 }

\title{\bf Mass-Deformed  Bagger-Lambert Theory
and its BPS Objects  }

\author{Kazuo Hosomichi, Ki-Myeong Lee,
and Sungjay Lee  \\
Korea Institute for Advanced Study,  Seoul 130-012, Korea
\\
{Email: hosomiti@kias.re.kr, klee@kias.re.kr, sjlee@kias.re.kr }}

\abstract{We find a sixteen supersymmetric  mass-deformed
Bagger-Lambert theory with $SO(4)\times SO(4)$ global R-symmetry.
The R-charge plays the `non-central' term in the superalgebra.
 This theory has one symmetric vacuum and two
in-equivalent broken sectors of vacua. Each sector of the broken
symmetry has the $SO(4)$ geometry. We find the 1/2 BPS domain
walls connecting the symmetric phase and any broken phase, and 1/4
BPS supertube-like objects, which may appear as anyonic q-balls in
the symmetric phase or vortices in the broken phase. We also discuss
mass deformations which reduce the number of supersymmetries. }

\begin{document}

\section{Introduction and Conclusion}

Recently there has been a spur of activities  on the possible
superconformal field theories for  the multiple M2 branes. The
so-called Bagger-Lambert
theory~\cite{Bagger:2006sk,Bagger:2007jr,Bagger:2007vi} has   16
supersymmetries and  $SO(8)$ global symmetry. However the
gauge matter coupling and the Chern-Simons
term are not standard but given by the three-product structures of
the $SO(4)$ group acting on vectors.  (See also the work by
Gustavsson~\cite{Gustavsson:2007vu}.)  Various  aspects of
this theory   have been
explored~\cite{Mukhi:2008ux,Bandres:2008vf,VanRaamsdonk:2008ft}.
Especially some detailed analyses  of the vacuum structure were done
to argue that this theory is the theory of two M2 branes on
M-theory orbifolds~\cite{Lambert:2008et,Distler:2008mk}.

In this work we find a mass deformation of the Bagger-Lambert
theory without breaking any supersymmetry. We find this mass
deformed theory is one example of the 3 dimensional supersymmetric field
theory with the so-called `non-central' term whose superalgebra
has been studied before~\cite{Nahm:1977tg,Lin:2005nh}.  The global R-charge
is  now $SO(4)\times SO(4)$ and plays the non-central term in the superalgebra.
We investigate the
vacuum structure and find  1/2 BPS and 1/4 BPS domain walls and
1/4 BPS localized solitons. This 1/4 BPS localized solitons are
basically supertubes which  appear  as q-balls in the symmetric
phase and vortices in the broken phases.

It has been known for sometime that the n-product object is
closely related to  a cross product of $n$  $(n+1)$ dimensional  vectors
to  one  $(n+1)$ dimensional vector. The three-product object used by  Bagger
and Lambert is realized as the  cross product
of three 4 dimensional vectors or $SO(4)$.  Since their work, there have been several attempts to
extend their structure~\cite{Berman:2008be,Morozov:2008cb,Gran:2008vi,Ho:2008bn,Bergshoeff:2008cz}.
However, any concrete realization remains to be seen.

While the ordinary Yang-Mills-Chern-Simons theory can have only
six supersymmetries~\cite{Kao:1992ig}, Lin and Maldacena have  found
a family of  eight supersymmetric Yang-Mills-Chern-Simons theories with
`noncentral' terms~\cite{Lin:2005nh}. This eight supersymmetric theory is
possible with arbitrary gauge group. It turns out to have
rich structures and allows  to include matter in other
representations as studied in~\cite{SJLEE}.  However,  we are not aware of any
explicit construction of the sixteen supersymmetric theory
with `non-central' term except the mass-deformed Bagger-Lambert theory studied here.

%%%%%%%%%%%%

The theory with mass deformation has one symmetric vacuum and two broken phases where
one of the global $SO(4)$ and the gauge $SO(4)$ are broken to the diagonal $SO(4)$. There is no massless
particle in each phase.   One can imagine the BPS domain walls connecting different sectors
of vacua. Indeed there are 1/2 BPS domain walls connecting the symmetric phase and the broken
phase, however, there is no BPS object connecting two broken phases. However we find
a non-BPS   domain   configuration explicitly as it satisfy a `fake' BPS equation.

The potential term suggests that there are attractive interactions between some particles.
One naturally expects some-sort of q-balls carrying R-charges. Similary, one may expect that
there  may be topological or non-topological vortices in the broken phase as the vacuum
manifold has $\pi_1(SO(4))=Z_2$. Indeed we find there are 1/4 BPS q-balls in the symmetric phase
and 1/4 BPS vortices in the broken phase. For a certain ansatz, the corresponding 1/4 BPS equations
become that of the abelian Chern-Simons theory~\cite{Hong:1990yh,Jackiw:1990aw,Jackiw:1990pr}, and so the previous analysis of the solitons is carried over to our case.   From the study of the 1/4 BPS domain walls which carry
the R-charges, one can see that the large-charge limit of the q-balls and vortices should be like
many other field theoretic supertubes~\cite{Kim:2006ee}. In this case both interior and exterior
of the soliton would be vacua and the boundary will be a domain wall carrying both R-charges and
linear momentum. Since the supertube has been proposed in Ref.~\cite{Mateos:2001qs}, there has been enormous work done on this subject but without any direct relevance to this work.

While the theory is parity even, one suspect the Chern-Simons term may still play a role in
the dynamics, leading to anyons of fractional spin and statistics. Indeed, we find the 1/4 BPS localized solitons carry fractional spin in the symmetric phase. It would be interesting to find
 the role  of the fractional spin and statistics in the mass deformed theory as the possible theory of two M2 branes on orbifolds.

%%%%%%%%%%%%%%%%%

The geometry of type IIB string theory with 16 supersymmetries and
with $S^1\times SO(4)\times SO(4)$ symmetries have been described by the droplets
of an incompressible fluid~\cite{Lin:2004nb}.  The theory of fermion droplet
on a cylinder with fermi sea is argued to describe the 16-supersymmetric
theory of M2 branes with a mass deformation whose vacuum structure describes the M2 branes
polarized into M5 branes wrapping two possible $S^3$'s~\cite{Bena:2000be,Pope:2003jp,Bena:2004jw,Lin:2005nh}.    The mass-deformed Bagger-Lambert theory may realize this picture.   The theory of the fermion droplet on a torus
is supposed to be also massive 16 supersymmetric theories, but is not written explicitly yet.     The dimensional reduction of the superalgebra to $1+1$ dimensions leads
to the linearly realized supersymmetries on the light-cone world sheet of a string moving in
the maximally supersymmetric type IIB plane wave~\cite{Blau:2001ne}. The investigation of these
pictures may lead to more explicit realization of    16 supersymmetric theories in 2+1 dimensions.

%%%%%%%%%%%%%%

The plan of this work is as follows. The mass-deformed theory is
introduced in Sec.2,  The vacuum structure  and the superalgebra
is given. In Sec.3, the 1/2  domain wall connecting the symmetric
phase to a broken phase is discussed. In Sec.4, we find the 1/4
BPS q-balls, vortices and supertubes. In Appendix, we find   mass-deformations
which reduce  the number of supersymmetries.

(While this paper is written, a massive deformation of the
Bagger-Lambert theory has been also proposed in
Ref.~\cite{Gomis:2008cv}.)

\section{Mass Deformed Theory with $SO(4)\times SO(4)$ Global Symmetry.}

The Bagger-Lambert theory has the $SO(4)$ bosonic variables
$X^a_I$, where $a=1,2,3,4$ for the $SO(4)$ index and $I=3,4,...10$
for $SO(8)$ index, and the eight independent component spinor
field $\Psi^a$. The gauge field is $A_\mu^{ab}$ which is
anti-symmetric tensor of $SO(4)$ gauge symmetry. The spacetime
signature is $(-++)$. We choose the 11 dimensional Gamma matrices which
are 32-by-32 matrices such that $\Gamma^0$ is antisymmetric real
matrix and the rest are symmetric real matrices. We choose the
convention for the spinor parameter so that
\be  \Gamma^{012}\Psi=-\Psi \ , \ \Gamma_{34..\bar{10}}\Psi=\Psi\  , \ \ \Gamma^{012}\epsilon =\epsilon \ , \
\Gamma_{34...9\bar{10}}\epsilon=-\epsilon\   . \ee
The gauge symmetry
is
 \be \delta X^a_I = f^{abcd} \Lambda^{cd} X^d_I \equiv\tilde{\Lambda}^{ab}X^b_I \ , \ee
 and the covariant derivative is
\be
D_\mu X^a_I = \partial_\mu X^a +f^{abcd} A_\mu^{cd} X^b_I \equiv
\partial_\mu X^a_I + \tilde{A}_\mu^{ab}X^b_I \ , \ee
with the local gauge transformation given as
\be \delta \tilde{A}_\mu = -D_\mu \tilde{\Lambda}_\mu = -(\partial_\mu \tilde{\Lambda}
+ \tilde{A}^{ac}_\mu \tilde{\Lambda}^{cb}  - \tilde{\Lambda}^{ac}\tilde{A}^{cb}_\mu ) \ .\ee
The field strength  given by  $[D_\mu, D_\nu]X^a= \tilde{F}_{\mu\nu}^{ab} X^b$ leads to
\be \tilde{F}^{ab}_{\mu\nu} = \partial_\mu \tilde{A}_\nu^{ab}  -\partial_\mu
 \tilde{A}_\nu^{ab} + \tilde{A}^{ac}_\mu
\tilde{A}^{cb}_\nu -
\tilde{A}^{ac}_\nu \tilde{A}^{cb}_\mu \ .\ee
As the three product structure has been realized only for the $SO(4)$ group, our $f^{abcd}=\epsilon^{abcd}$.

The supersymmetric Lagrangian for the Bagger-Lambert theory is
\ba {\cal L} &=&-\frac12  D_\mu X_I^a D^\mu X_I^a + \frac{i}{2}  \bar{\Psi}^a
 \Gamma^\mu D_\mu \Psi^a   \nonumber \\
& & - \frac{i}{4\kappa} f^{abcd}  \bar{\Psi}^a   \Gamma_{IJ}\Psi^b X^c_I X^d_J
   -\frac{1}{12\kappa^2}\sum_{a,I,J,K}\Big(f^{abcd}X^b_{I} X^c_J X^d_{K} \Big)^2\  \nonumber \\
& &  +\frac{\kappa}{2} \epsilon^{\mu\nu\rho} \Big( f^{abcd}
A_\mu^{ab}
\partial_\nu A_\rho^{cd}+ \frac{2}{3}f^{acde}f^{bcfg} A_\mu^{ab}
A_\nu^{de} A^{fg}_\rho \Big) \ , \ea
where $\bar{\Psi}^a = \Psi^{a\dagger} \Gamma^0 $.
We propose the mass-deformation of the theory to be
\be {\cal L}_m= -\frac{m^2}{2} ( X_I^a)^2 - \frac{i}{2}
m\bar{\Psi}^a\Gamma_{3456}\Psi^a   + \frac{4m}{\kappa} f^{abcd}
(X_3^a X_4^b X_5^c X_6^d +X_7^a X_8^b X_9^c X_{10}^d) \ . \label{mass1} \ee
The bosonic part of the mass deformation for $3,4,5,6$ was studied in Ref.~\cite{Bena:2000be,Bagger:2007jr}.
In this work the broken phase of the theory is also identified as the M2 branes blowing up to M5 branes wrapping $S^3$.
The supersymmetric transformation of the fields also get deformed
as
\ba && \delta X_I^a = i\bar{\epsilon}\Gamma_I\Psi^a \ ,  \\
&& \delta \Psi^a = (  \Gamma^\mu    D_\mu  + m \Gamma_{3456})X^a_I
\Gamma_I\epsilon  + \frac{1}{6\kappa}
\Gamma_{IJK}\epsilon\  f^{abcd} X_I^b X_J^c X_K^d  \ , \\
&& \delta A_\mu^{ab} = -
 \frac{i}{2\kappa}   \bar{\epsilon}\Gamma_\mu \Gamma_I (\Psi^aX^b_I-\Psi^b X^a_I)
 \ . \ea
The Gauss law
constraint arising from $\delta A_0^{ab}$ is
\be \kappa \tilde{F}^{ab}_{12} - f^{abcd} X^c_I  D_0 X^d_I
+ \frac{i}{2}  f^{abcd}\bar{\Psi}^c \Gamma^0  \Psi^d = 0 \ . \ee
The quantization of the Chern-Simons coefficient is
\be 2\pi \kappa = n \ ,  \ee
with integer $n$. The original global $SO(8)$ symmetry is broken to $SO(4)\times
SO(4)$ symmetries.  Among the conserved charge for the original
$SO(8)$ symmetry
\be R_{IJ} = \int d^2x \Big(  X^a_I D_0 X_J^a- X^a_J D_0 X_I^a
 + \frac{i}{2} \bar{\Psi}^a\Gamma^0\Gamma_{IJ}\Psi^a ) \ , \ee
only those   $SO(4)\times SO(4)$,  each of which rotates $3,4,5,6$ and $7,8,9,10$  respectively,  will be preserved.

The bosonic potential of the theory  can be written as
\be U(X) = \frac{1}{12\kappa^2} \sum_{a, J,K,L} \Big(f^{abcd}
X_J^b X_K^c X_L^d
 - \kappa m
\delta^{3456}_{IJKL}X^a_I - \kappa m \delta^{789\bar{10}
}_{IJKL}X_I^a \Big)^2 \ , \ee
where $\delta^{3456}_{IJKL}= \epsilon_{IJKL}$ for $I,J,K,L\in \{
3,4,5,6\}$ and vanishes for other combinations, and
$\delta^{78910}_{IJKL}$ is defined similar manner. There exist
three  independent sectors  of vacua in the theory. The symmetric
phase where $X_I^a =0 $ for all $I$ and two broken phases where
\ba     ({\bf I}) &&  \ \   X_I^a=\sqrt{|\kappa m|} E_I^a \ , \ I=3,4,5,6  \ ,  \\
({\bf II}) &&  \ \  X_I^a = \sqrt{|\kappa m|} E_I^a\ , \
I=7,8,9,\bar{10} \ . \ea
The $\{ E_3^a, E^b_4, E^c_5, E^d_7 \} $ and $\{ E_7^a, E_8^b,
E_9^c, E_{\bar{10}}^d \} $ are two sets of four dimensional
orthonormal frame of ${\bf R}^4$ such that
\ba &&  E_I^a E_J^a= \delta_{IJ}\ , \ E^a_I E^b_I =\delta^{ab} \ , \nonumber \\
&& f^{abcd}E^a_I E^b_J E^c_K E^d_L = {\rm sgn}(\kappa m)(  \delta^{3456}_{IJKL}
+ \delta^{789\bar{10}}_{IJKL}  ) \ .
\ea
%.
These vacua has the zero energy and are fully supersymmetric. In
one of the broken phase, the $SO(4)$ gauge group and one $SO(4)$
of the global group gets locked together into a single $SO(4)$
symmetry.  The vacuum structure of each broken sector is the manifold of the
$SO(4)$ gauge group, which is six-dimensional.  As the first
homotopy group of the $SO(4)$ manifold is $Z_2$, one may expect
$Z_2$ vortices in the broken phase. Of course we can mod out the
global gauge symmetry to get a point for each vacuum.

The superalgebra can be checked easily by the communtation
relation
\be [\delta_\eta, \delta_\epsilon] X^a_I = 2a^\mu \partial_\mu
X^a_I + 2 f^{abcd}\Lambda^{cd} X^b_I + 2m \sum_{ J} S_{IJ} X_J^a, \ee
where the parameters for the translation, gauge transformation,
and the global $SO(4)$ rotations are given respectively as
\ba && a^\mu = i\bar{\eta}\Gamma^\mu \epsilon \ , \nonumber \\
&& \Lambda^{cd} =  i\bar{\epsilon}\Gamma^\mu\epsilon A_\mu^{cd}
-\frac{1}{2\kappa}i\bar{\eta}\Gamma_{JK}\epsilon X_J^c X^d_K  \ ,\nonumber \\
&& S_{IJ}= ( \delta^{3456}_{IJKL}+ \delta^{789\bar{10}}_{IJKL} )
i\bar{\eta}\Gamma_{KL}\epsilon \ . \ea
As the supercharge is not invariant under the global $SO(4)\times
SO(4)$, their charge is called as the `non-central' term. Indeed
this noncentral terms are essential in reducing the supersymmetric
representation of the massive particles in each phase.

The spectrum of the particles in the symmetric phase is trivial in the week coupling
constant limit $n>>1$. There are
$32$ scalar particles of mass $m$ for the field $X_I^a$ and $32$
fermionc particles of mass $m$ for the field $\Psi$. In the broken phase $(I)$, the Higgs mechanism
leads to massive vector bosons, and the global rotation $SO(4)_1$ rotating $3,4,5,6$ variables and the gauge
symmetry gets soldered together. In the broken phase, all particles have mass $2m$.
There are six massive vector bosons, 32  massive fermions and 26 scalar bosons.  As the theory is
parity even, the spin content is $(3,16,26,26,3)$ number of elementary particles for the spin $(1,1/2,0,-1/2,-1)$.
Due to the Chern-Simons term, elementary particles in the symmetric phase may carry fractional statistics.

\section{Domain Walls}

Let us first consider the domain wall connecting the symmetric
vacuum to the asymmetric vacuum of the first broken phase $(I)$.
As we assume that $X_{7,8,9,\bar{10}}=0$, we introduce a
`superpotential'
\be W = \frac{m}{2} \sum_{I=3,4,5,6} (X_I^a)^2 - \frac{1}{\kappa}
f^{abcd}X^a_3 X^b_4 X^c_5 X^d_6  \ ,\ee
and then the bosonic potential becomes $U= |W^a_I|^2/2 $ with
\be W^a_I= \frac{\partial W}{\partial X_I^a} = mX^a_I -
\frac{1}{6\kappa} f^{abcd} \delta^{3456}_{IJKL} X^b_J X^c_K X^d_L \ .
\ee
The energy density along the wall becomes
\be E=\frac{1}{2} \int dy \big(  D_y X_I^a -\beta W_I^a \big)^2 \
+ \beta {\cal T}  \ , \ee
where $y=x^2$ and $\beta=\pm 1$ and
\be {\cal T} =  \int dy  \partial_y W  |\kappa m^2|\ .  \ee
Thus the tension of the domain wall becomes
\be |{\cal T}|= |\kappa m^2| \ .  \ee
 The BPS equation is
\be D_y X_I^a -\beta W_I^a= 0  \ . \ee
We assume $\kappa, m >0$ for the convenience and use the ansatz

\be A_2^{ab}=0\ , \  X^a_I= \sqrt{ \kappa m} f(y) {\rm diag}(1,1,1,1)\ , \ee
where the row indices are $I$ and the column indices are $a$. The
BPS equation becomes
\be \partial_y f-\beta m (f-f^3) = 0\ ,  \ee
whose solution is
\be f= \sqrt{\frac{1+\beta \tanh (my)}{2}} \ .\ee
One can also consider the simple generalization of the above
domain wall which connects the symmetric phase to the second
broken vacua  $({\bf II})$.  Clearly  both domain wall solutions
are  1/2 BPS configuration. The above configuration is 1/2 BPS as
the invariant condition for $\delta \Psi$ is
\be \Gamma_{23456}\epsilon=\epsilon\ . \ee

One can wonder whether there is any domain wall configuration which
interpolates two broken phases. Indeed we can find an
interpolating configurations which satisfies the first order
equations, but not BPS as the supersymmetric condition is not
satisfied. We combine the above ansatz and
\be X^a_I = g(y)\sqrt{\kappa m } {\rm diag}(1,1,1,1)\ , \
I=7,8,9,\bar{10} \ . \ee
We impose the wrong BPS condition on the epsilon
$\Gamma_{23456}\epsilon=\Gamma_{2789\bar{10}}\epsilon=\epsilon$ and
$(\Gamma_{37}+\Gamma_{48}+\Gamma_{59}+\Gamma_{6\bar{10}})\epsilon=0$,
to get the wrong BPS equations
\be  f' + m(f^3-f + fg^2)=0 \ ,   -  g'+m(g^3-g + gf^2)=0 \ . \ee
whose 1-parameter family of the solutions are
\be f^2 = \frac{(1+a)(1+\tanh my)^2}{4(1+ a \tanh^2my)}\ , \ g^2 = \frac{(1+a)(1-\tanh my)^2}{4(1+ a \tanh^2my)} \ .\ee
We expect  that two walls are repulsive and so the above configuration will not remain static in time.

\section{Q-Balls, Vortices, and Supertubes}

 As the $3,4,5,6$ particles can attract each other and condense
into one of the broken phases, we expect that there can be a lump
of these $3,4,5,6$ particles, or equally lumps of the
$7,8,9,\bar{10}$ particles. Due to the noncentral terms, one may
have a BPS q-balls or nontopological solitons, carrying R-charges. Indeed we will see
here that there exist  q-balls in the symmetric phase and vortices
in the broken phase, whose equations are identical but with
different boundary conditions.

We  expect that spatial coordinates and  $X_{3,4,5,6}$ get
involved and $X_{7,8,9,\bar{10}}=0$. We introduce thus a  1/4 BPS
condition on the spinor parameter such as
\be \Gamma_{1234}\epsilon=-\alpha\epsilon\ ,\
\Gamma_{1256}\epsilon=-\beta \epsilon \ , \ee
where $\alpha, \beta =\pm 1$. This implies that
$\Gamma_{3456}\epsilon=-\alpha\beta \epsilon$.  The variation of the spinor field
vanishes if the time-derivative parts and the potential parts are
matched so that
\ba
&& D_0X_3^a+ \beta W_4^a=0 \ , \ D_0X_4^a- \beta W_3^a=0 \ , \nonumber\\
&& D_0 X_5^a+ \alpha W_6^a=0 \ , \ D_0X_6^a-\alpha W_5^a= 0 \ , \ea
and the spatial derivatives are matched so that
 \ba
&& D_1 X_3^a-\alpha D_2X_4^a=0 \ , \ D_2X_3^a+\alpha D_1X_4^a=0\ , \nonumber  \\
&& D_1X_5^a -\beta D_2 X_6^a=0 \ ,  \ D_2 X_5^a+ \beta D_1 X_6^a
=0 \ .\label{vortexbps}\ea
We can go over the energy and find a BPS bound on the energy
\be {\cal E} \ge  m(\beta R_{34} +\alpha R_{56})= m(
|R_{34}|+|R_{56}|)\ , \ee
once we use the Gauss law and the signs $\alpha$ and $\beta$ is
chosen so that the equality in the right side holds. This bound is
saturated by the configurations which satisfy the equations
(\ref{vortexbps}). For the $\alpha,\beta=1$, the total $R$ charge
for the BPS configurations becomes
\be R_{34} +R_{56} = m\sum_{I=3,4,5,6}(X_I^a)^2 -\frac{4}{\kappa}
f^{abcd} X^a_3 X^b_4 X^c_5 X^d_6 \ ,  \ee
which vanishes for the symmetric phase and the broken phase.

To understand the 1/4 BPS configuration, let us use the the ansatz
 \be
 X^a_I= \left(\begin{array}{cccc}  \phi_1 & -\phi_2 & 0 & 0 \\
  \phi_2 & \phi_1 & 0 & 0\\
 0 & 0& \phi_1 & -\phi_2 \\
0 & 0 & \phi_2  & \phi_1  \end{array}\right)\ ,
\  \tilde{A}^{ab}_\mu \equiv \left( \begin{array}{cccc}
0 & -A_\mu & 0 & 0 \\
A_\mu & 0 & 0 & 0 \\
0 & 0& 0 & -A_\mu \\
0 & 0 & A_\mu & 0 \end{array}\right) \ ,\ee
where the row indices for $X^a_I$ is $I$ and the column indices
$a$. Introducing a  complex scalar $\phi=\phi_1+i\phi_2$, the above 1/4 BPS equations and the Gauss law  reduce to
 \be (D_1 +iD_2)\phi=0 \ , \
  F_{12} = \frac{2}{\kappa^2} |\phi|^2  (m\kappa -|\phi|^2) \ , \ee
where $D_i\phi = \partial_i \phi -i A_i \phi$ and
$F_{12}=\partial_1 A_2-\partial_2 A_1$. This is the BPS equation
for q-balls and vortices in the selfdual Chern-Simons Higgs theory
with fractional spin. Indeed the angular momentum for the BPS
configuration is
\ba J_{12} &=& -\int d^2x (x_1  D_0X_I^a D_2 X_I^a -x_2 D_0X_I^a
D_1
X_I^a  \nonumber) \\
&=& \int d^2 x x_i \partial_i W \ .\ea
Q-balls can have vortices in interior region.
Especially one can show that the angular momentum for q-balls is
\be  J_{12}=  \frac{(R_{34}+R_{56})^2}{4\pi \kappa}\ , \ee
which shows that   q-balls can carry fractional angular
momentum~\cite{Jackiw:1990pr}. (For
the vortices, there woul be  a sign-flip.) This
makes the details of the representation of the superalgebra interesting.

As in Ref~\cite{Kim:2006ee}, the q-balls and vortices in the large charge
limit can be regarded as  supertubes in the symmetric phase or the
broken phase. For this interpretation, it was important that the
domain wall interpolating the symmetric phase and one of the
broken phase attracts the R-charges to the wall, forming a
composite objects and a novel BPS bound on the domain wall was
found. We show that this is true again for the theory under
consideration.

We can consider additional R-charges on the domain
walls. Let us again impose the supersymmetric conditions
\be e^{\theta \Gamma_{134}}\Gamma_{1234}\epsilon=-\alpha \epsilon\
, \ e^{\theta \Gamma_{134}}\Gamma_{1256} \epsilon=-\beta\epsilon \ ,
\ee
where $\alpha,\beta=\pm 1$. This leads again to
$\Gamma_{3456}\epsilon=-\alpha\beta \epsilon$. Here the angle
$\theta$ is arbitray and will be fixed for a BPS configurations.
Note that $\theta=0$ is the BPS condition for   q-balls and
vortices and $\theta=\pi/2$ is that for the domain wall with R-charges.
  The supersymmetric BPS equations for the 1/4 BPS
domain wall carrying R-charges are made of eight equations as follows:
\ba
&& D_0 X^a_3  +\beta    W^a_4 \cos\theta   - \alpha D_1 X_3^a \sin\theta =0\ , \nonumber  \\
&& D_0 X_4^a  -\beta   W^a_3\cos\theta -\alpha  D_1 X_4^a\sin\theta  =0\ , \nonumber  \\
&& D_0 X_5^a  +\alpha   W^a_6\cos\theta -\alpha   D_1 X_5^a\sin\theta  =0\ , \nonumber  \\
&& D_0 X_6^a  -\alpha      W^a_5\cos\theta  -\alpha  D_1 X_6^a
\sin\theta =0\ , \nonumber \\
&& D_2 X_3^a  +\alpha D_1 X_4^a \cos\theta   -\beta W_3^a\sin\theta  =0\ , \nonumber\\
&&    D_2 X_4^a  - \alpha   D_1X_3^a\cos\theta  -\beta W_4^a\sin\theta =0\ , \nonumber  \\
&&  D_2 X_5^a  +\beta  D_1X_6^a\cos\theta - \beta W_5^a\sin\theta =0 \ , \nonumber  \\
&& D_2 X_6^a -\beta  D_1 X_5^a\cos\theta -\beta W_6^a\sin\theta=0\ .
\ea
Once these equations are satisfied,  the energy density integrated
along $x^2$ direction becomes
\be {\cal E}= m ( -\beta R_{34} -\alpha R_{56})\cos\theta +  (
\alpha {\cal P} +\beta  {\cal T})\sin\theta \ ,  \ee
where the linear momentum density and the signed wall tension are,
respectively,
\ba && {\cal P} = \int dy  D_0 X^a_I D_1 X^a\ , \\
&& {\cal T}=  \int dy  \partial_y W  \ .
\ea
As argued before, $|{\cal T}|= \kappa m^2$. The BPS bound would be
 \be {\cal E} \ge \sqrt{ m^2(\beta R_{34} +\alpha R_{56})^2+
 (\beta {\cal T}+\alpha {\cal P})^2} \ee
for all $\alpha,\beta$. Indeed when $|{\cal T}|=|{\cal P}|$, we
can choose $\alpha $ and $\beta$ so that $\beta{\cal
T}+\alpha{\cal P}=0$, so that the energy density along the wall is
just given by the R-charges, and there is a momentum flow along
the wall which is given the wall tension. This is exactly the
supertube condition. In the supertube limit ($\sin\theta=0$), the above BPS equations
become those for the 1/4 BPS q-balls and vortices.

\noindent{\bf Acknowledgment}

K.M.L. is supported in part by  the KOSEF SRC Program through
CQUeST at Sogang University, KRF Grant No.
KRF-2005-070-C00030, and the KRF National Scholar program.

\appendix

\section{ Mass Deformation  with Less Supersymmetries}

One can introduce  mass deformations which break supersymmetry partially.
Instead of the single mass term in (\ref{mass1}), we introduce three mass parameters
 $m,m',m''$ so that the fermionic mass term becomes
\be {\cal L}_{fm}= -  \frac{i}{2} \bar{\Psi}^a ( m\Gamma_{3456} + m'\Gamma_{3478}+ m''\Gamma_{349\bar{10}} )
\Psi^a.\ee
Again, only the fermionic supersymmetric transformation gets modified by the additional expression
\be \delta_m\Psi^a = (m\Gamma_{3456}+ m'\Gamma_{3478}+m''\Gamma_{349\bar{10}} )\Gamma_I \epsilon \ X^a_I \ . \ee
We impose  three constraints on the $\epsilon$ parameter,
\begin{eqnarray} 
\Gamma_{5678}\epsilon &=& - \alpha \epsilon \ ,  \nonumber \\ \Gamma_{569\bar{10}}\epsilon &=& - \beta \epsilon \ , \nonumber \\ \Gamma_{789\bar{10}}\epsilon &=& -\alpha\beta \epsilon .
\end{eqnarray}
Only two of them are independent and so the number of  supersymmetries is reduced to four. (If $m''=0$, the number would be eight.) We then introduce following bosonic interactions $\CL_{\rm deformed} = \CL_{\rm mass} + \CL_{\rm quartic}$ for supersymmetric completion
\begin{eqnarray}
  \CL_{\rm mass} &=& -\frac{1}{2} m^2_{IJ} X_I^a X_J^a, \nonumber \\ \CL_{\rm quartic} &=& \frac{4m}{\kappa} f_{abcd} (X^a_3 X^b_4 X^c_5 X^d_6 + X^a_7 X^b_8 X^c_9 X^d_{\bar{10}}) \nonumber \\ 
  && \hspace{0.2cm} + \frac{4m'}{\kappa} f_{abcd} (X^a_3 X^b_4 X^c_7 X^d_8 + X^a_5 X^b_6 X^c_9 X^d_{\bar{10}}) \nonumber \\ && \hspace{0.2cm} + \frac{4m''}{\kappa} f_{abcd} (X^a_3 X^b_4 X^c_9 X^d_{\bar{10}} + X^a_5 X^b_6 X^c_7 X^d_8),
\end{eqnarray}
where the mass matrices for the bosonic fields $X_I^a$ are diagonal and given by
\ba && (m^2)_{33}=(m^2)_{44}= (m+\alpha m'+ \beta m'')^2\ , \nonumber  \\
&& (m^2)_{55}=(m^2)_{66}=(m -\alpha m'- \beta m'')^2 \ , \nonumber \\
&&  (m^2)_{77}=(m^2)_{88}=(m- \alpha m'+ \beta m'')^2 \ ,\nonumber \\
&&  (m^2)_{99}=(m^2)_{\bar{10}\bar{10}}=(m+\alpha m'-\beta m'')^2 . \ea 
Our approach for the mass deformation seems to have some analogy with the mass deformation
of the supergravity solution for $AdS_4\times S^7$ geometry in Ref.~\cite{Bena:2000be}.

\end{document}